\documentclass{elsart}

\usepackage{natbib}

 \usepackage{epsfig}

\usepackage{amssymb}

\begin{document}
\newcommand{\lsun}{\ensuremath{\,{\rm L}_\odot}} 
\newcommand{\ion}[2]{\mbox{#1\,{\scriptsize #2}}}
\newcommand{\mdot}{\mbox{$\dot{M}$}}
\newcommand{\Teff}{\mbox{$T_\mathrm{eff}$}}
\newcommand{\msun}{\ensuremath{\, {\rm M}_\odot}}
\newcommand{\logg}{\mbox{$\log g$}}
\newcommand{\exo}{\mbox{EXO\,0748$-$676}}

\begin{frontmatter}

\title{Non-LTE Models for Neutron Star Atmospheres and Supernova-Fallback Disks}

\author[label1]{K. Werner},
\ead{werner@astro.uni-tuebingen.de}
\author[label1]{T. Nagel},
\author[label1]{T. Rauch}, and 
\author[label1,label2]{V. Suleimanov}

\address[label1]{Institut f\"ur Astronomie und Astrophysik, Universit\"at T\"ubingen, Sand 1, 72076 T\"ubingen, Germany}
\address[label2]{Kazan State University, Kremlevskaja Str., 18, Kazan 420008, Russia}

\begin{abstract}
We describe our recent progress in modeling supernova-fallback disks and
neutron star (NS) atmospheres.

We present a first detailed spectrum synthesis calculation of a
SN-fallback disk composed of iron. We assume a thin disk
with a radial structure described by the $\alpha$-disk model.  The
vertical structure and emission spectrum are computed self-consistently
by solving the structure equations simultaneously with the radiation
transfer equations under non-LTE conditions. We describe the properties of a
specific disk model and discuss various effects on the emergent UV/optical
spectrum.

We investigate Compton scattering effects on the thermal spectrum of
NSs. In addition, we constructed a new generation of metal
line-blanketed non-LTE model atmospheres for NSs. It is compared to
X-ray burst spectra of \exo. It is possible that the gravitational
redshift, deduced from absorption lines, is lower ($z$=0.24) than hitherto
assumed ($z$=0.35). Accordingly, this would result in NS 
mass and radius lower limits of $M$\,$\geq$1.63~\msun\ and $R$\,$\geq$13.8~km. 
\end{abstract}


\end{frontmatter}

\section{Introduction}

We report on our recent progress in radiation transfer modeling for
supernova-fallback disks and non-magnetic neutron star (NS) atmospheres. Disk
modeling is described in Sect.~\ref{disks}. We then present effects of Compton
scattering on the thermal spectra of NS atmospheres (Sect.~\ref{compton}). In
Sect.~\ref{nsmodels} we describe the construction of new line-blanketed NLTE
atmosphere models for NSs. These models are applied to X-ray burst spectra of
\exo\ (Sect.~\ref{exo}).

\section{NLTE models for SN fallback disks}\label{disks}

Anomalous X-ray pulsars (AXPs) are slowly rotating young isolated NSs. Their
X-ray luminosities greatly exceed the rates of rotational energy loss.  It is
now generally believed that AXPs are magnetars and that their X-ray luminosity
is powered by magnetic energy. As an alternative explanation the X-ray emission
was attributed to accretion from a disk that is made up of supernova-fallback
material \citep{Par95}. It is still a matter of debate whether the disk model is
appropriate. For example,  the discovery of optical pulsations in 4U\,0142+61
with the same period like the X-ray pulsations \citep{Ker02} appears to be
a strong argument against the disk model. It was argued that reprocessing of the
pulsed NS X-ray emission in the disk cannot explain the high optical pulsed
fraction, because disk radiation would be dominated by viscous dissipation and
not by reprocessed NS irradiation \citep{Ker02}.  In contrast, \citet{Ert04}
showed that the optical pulsations can  be explained either by the magnetar
outer gap model or by the disk-star dynamo model.  A spectral break in the
optical spectrum of 4U\,0142+61 was discovered by \citet{Hul04}  and also taken
as an argument against the disk model.  The recent discovery of mid-IR emission
from this AXP \citep{Wan06}, however, has strongly rekindled the interest in
studies of fallback-disk emission properties. The emission was attributed to a
cool, passive dust debris disk, but it was shown by \citet{Ert06}  that it can
be explained by an active, dissipating gas disk.

The emission from fallback disks was hitherto modeled with blackbody spectra. In
view of the importance of disk models for the quantitative interpretation of
observational data it is highly desirable to construct more realistic models by
detailed radiation-transfer calculations. We employ our computer code AcDc
\citep{Nag04}. For a detailed description of the calculation of fallback disk
models we refer to \citet{Wer06}. The radial disk structure is calculated
assuming a stationary, Keplerian, geometrically thin $\alpha$-disk.  For the
results reported here we selected the following model parameters. The NS mass is
1.4~\msun. The radii of the inner and outer disk edges are 2000 and 200\,000~km,
respectively. The accretion rate was set to $\dot{M}=3\cdot
10^{-9}$~\msun/yr, which is a relatively large value. It is motivated
by the fact, that this corresponds to the upper limit for the non-detection of UV
radiation from a possible disk around SN~1987A under the assumption
of an outer disk radius of 100\,000~km \citep{Gra05}. Our code allows
for the irradiation of the disk by the central source, however, the results
presented here are computed with zero incident intensity. 
The disk is represented by nine rings.  The vertical structure of each ring
is determined from the simultaneous solution of the radiation transfer equations
plus the structure equations (radiative and hydrostatic equilibrium). They also
consist of the non-LTE rate equations for the atomic population densities.  The
chemical composition of SN fallback material is not exactly known. 
It depends on the amount of mass that goes into the disk. A disk
with a small mass (say $\leq 0.001$~\msun) will be composed of silicon-burning
ash \citep{Men01}. For
simplicity we assume a pure-iron disk composition but also tested a composition
that represents silicon-burning ash. It contains iron (80\% mass fraction) as well as
silicon and sulfur (10\% each). Fig.~\ref{spectrum_entire_disk_rotation} displays the
overall disk spectrum. We summarize the model properties as follows:

\begin{figure}
\centering
  \includegraphics[width=0.7\columnwidth]{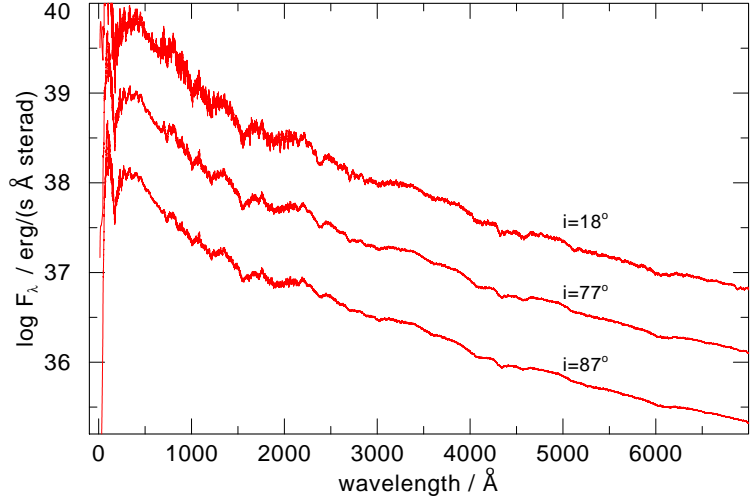}
\caption{Spectrum of a fallback-disk model seen under three
  inclination angles $i$. Broad iron-line blends are detectable even in the
  almost edge-on case.}
\label{spectrum_entire_disk_rotation}
\end{figure}

\begin{itemize}

\item Depending on the inclination, the disk flux can be a factor of two higher
or lower compared to a blackbody radiating disk at the same inclination.

\item Strong iron line blanketing causes broad ($>$100~\AA) spectral features
that could be detectable even from almost edge-on disks.

\item Limb darkening affects the overall disk spectrum (in addition to the
geometric foreshortening factor). Depending on inclination and spectral
band, the disk intensity varies up to a factor of three.

\item The overall disk spectrum is independent of the detailed chemical
composition as long as Fe is the dominant species. In particular, a pure-Fe
composition is spectroscopically indistinguishable from Si-burning ash.

\item The overall disk spectrum is hardly influenced by non-LTE effects,
however, equivalent widths of individual line blends can change by a factor of
two.

\end{itemize}

\begin{figure}
\centering
  \includegraphics[width=0.75\columnwidth]{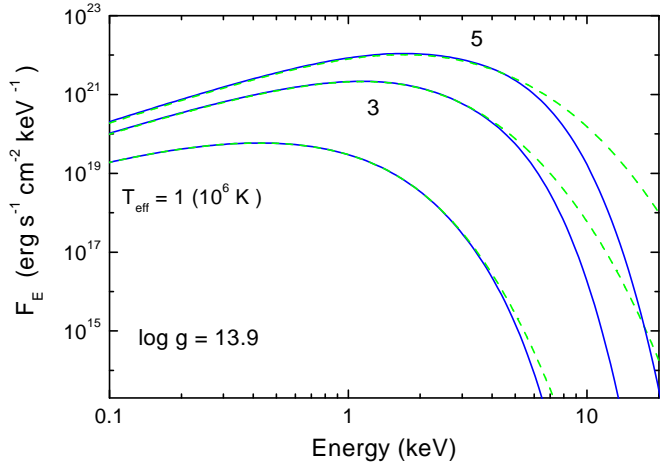}
\vspace{-1cm}
\caption{Emergent spectra of three pure-H models with different $T_{\rm
  eff}$ ($1, 3, {\rm and}~5 \cdot 10^6$~K) computed with Thomson (dashed lines) and
  Compton (full lines) electron scattering.}
\label{sul_fg1b}
\end{figure}

\section{Compton scattering in NS model atmospheres}\label{compton}

The nature of isolated NS (INS) surface layers is not exactly known. At some
conditions (depending on surface temperature, magnetic field strength and
chemical  composition) a surface can be solid, liquid, or have a plasma envelope
\citep{Lai97, Lai01}. In the last case the envelope can be considered as a NS
atmosphere, and  the structure and emergent spectrum of this atmosphere can be
computed by using stellar model  atmosphere methods. Such modeling was performed
by many groups and the model spectra were used to fit observed INS X-ray
spectra.  One of the important results of these works is that
spectra of light elements (hydrogen and helium) atmospheres with low magnetic
field are significantly harder than the corresponding blackbody spectra. These
elements are fully ionised in  atmospheres with $T_{\rm eff} \ge
10^6$~K. Therefore, the true opacity (mainly due to free-free transitions)
decreases with photon energy as $E^{-3}$. At high energies electron scattering
is larger than the true opacity and photons emitted deep in the  atmosphere
(where $T > T_{\rm eff}$) escape after few scatterings on electrons. In all 
previous work concerning INS, coherent (Thomson) electron scattering is
considered. As a result,  emergent spectra are very hard. But such a situation
is very favorable to change the photon energy due to Compton down-scattering.

It is well known that Compton down-scattering determines the shape of
emergent model spectra of hotter NS atmospheres with $T_{\rm eff} \sim 2 \cdot
10^7$ K close to the Eddington limit \citep{Pav91}. These model spectra
describe the  observed X-ray spectra of X-ray bursting NS in low-mass X-ray
binaries (LMXBs), and they are close to diluted blackbody spectra with a hardness
factor $f_c \sim$~1.5--1.9.   But models with Compton scattering taken into
account were not calculated for relatively cool atmospheres with $T_{\rm eff} <
10^7$~K. Therefore, at present time, the effect of Compton scattering on
emergent spectra of INS model atmospheres with $T_{\rm eff} < 5 \cdot 10^6$~K is
not well known.

We investigated the effect of Compton scattering on hot ($T_{\rm eff} > 10^6$~K)
INS atmospheres with weak magnetic fields ($B<10^8$~G). In order to compute LTE model
atmospheres in hydrostatic and radiative equilibrium we solve the radiation
transfer equation with the Kompaneets operator. We calculated a set of models
with $T_{\rm eff}$ in the range 1--5$ \cdot 10^6$~K, with two values of surface
gravity (\logg=13.9 and 14.3), and different chemical compositions
\citep{Sul07} and the results can be summarized as follows:

\begin{itemize}

\item Radiation spectra computed with Compton scattering are softer than those
computed without Compton scattering at high energies ($E>$~5~keV) for light
element (H or He) model atmospheres (Fig.~\ref{sul_fg1b}).

\item The Compton effect is most significant in H model atmospheres and models
with low surface gravity.

\item Compton scattering is less important in models with solar abundance of
heavy elements.

\item The emergent spectra of the hottest ($T_{\rm eff} >3 \cdot 10^6$~K) models
can be described by diluted blackbody spectra with hardness factors
$\sim$~1.6--1.9.

\end{itemize}

\section{Metal line blanketed NLTE models for NS atmospheres}\label{nsmodels}

\begin{figure}
\centering
  \includegraphics[width=\columnwidth]{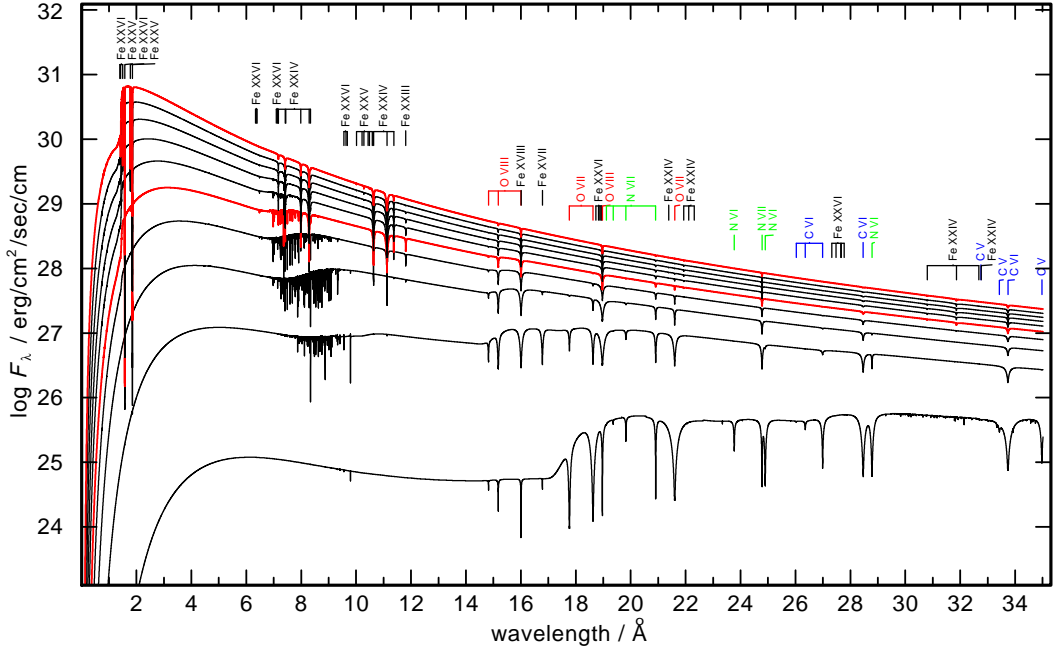}
\caption{Spectra of solar composition NLTE model atmospheres with
  increasing $T_{\rm eff}$ (1--10$\cdot 10^6$~K in steps of $10^6$~K) at \logg=14.39.}
\label{Fig4_mod}
\end{figure}

We are computing a new grid of model atmospheres for weakly magnetized
INSs. Since it covers a large \Teff-range up to high temperatures (1--10$\cdot
10^6$~K) it takes into account deviations from LTE for the first
time. \citet{We00}  have shown that NLTE effects can affect the temperature
structure of the outer photospheric layers. The models are line-blanketed and
calculated for different chemical compositions comprising H, He, C, N, O, and
iron (Rauch, Suleimanov, Werner, in prep.). We employ the {\sc TMAP} (T\"ubingen
Model Atmosphere Package)  codes \citep{We03} and account for opacities of
millions of iron lines in a statistical way \citep{Ra03}. The model atoms are
constructed using energy levels from  NIST and oscillator strengths and
photoionisation cross-sections calculated by the Opacity Project  (TIPTOPbase).
As an example, in Fig.~\ref{Fig4_mod} we display the X-ray flux spectra of
models with different \Teff, solar element composition, and \logg=14.39.

\section{Model comparison to X-ray burst spectra of \exo}\label{exo}

\citet{Co02} have announced the detection of absorption features in the burst
spectra of the LMXB \exo. The three most significant features
were identified as photospheric absorption lines, redshifted by $z$=0.35: A
feature at 13.00~\AA\ in the early-burst phases was assigned to the
\ion{Fe}{XXVI} n=2--3 transition at $\lambda_0$=9.5--9.7~\AA. In the late-burst
phase a feature appears at 13.75~\AA\ and it was assigned to the \ion{Fe}{XXV}
n=2--3 transition at $\lambda_0$=10.2~\AA. A double feature at 25.2/26.0~\AA,
also appearing in the late-burst phase, was interpreted as a broad self-reversed
line profile from \ion{O}{VIII}~Ly$_\alpha$ ($\lambda_0$=18.97~\AA).

\begin{figure}
\centering \includegraphics[width=1.0\columnwidth]{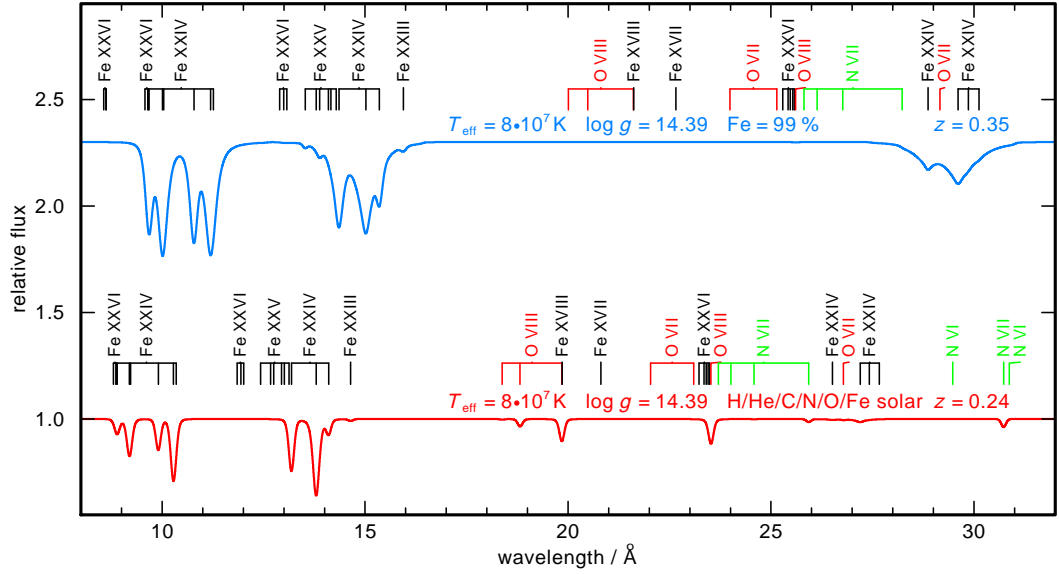}
\caption{Spectra from two models with $z$=0.35 (top) and
  $z$=0.24 (bottom). They suggest that the observed absorption lines in \exo\
  at $\lambda$=13.0/13.75~\AA\ could stem from \ion{Fe}{XXIV} at $z$=0.24 rather
  than from \ion{Fe}{XXVI}/ \ion{Fe}{XXVII} at $z$=0.35.}
\label{exomodels}
\end{figure}

We can use our models to check these line identifications. The upper curve in
Fig.~\ref{exomodels} is the spectrum of a Fe-dominated (99\%) model with
\Teff=$8\cdot 10^7$~K, \logg=14.39, redshifted by $z$=0.35. This Figure can be
directly compared to Fig.~1 in \citet{Co02}. It is obvious that the
aforementioned \ion{Fe}{XXVI}/ \ion{Fe}{XXVII} lines at (redshifted)
$\lambda$=13.0/13.75~\AA\ are much too weak in the model. In models with
different \Teff\ these lines are even weaker, so that the identification by
\citet{Co02} is not confirmed.

Other Fe lines in the models are much stronger and we suggest another possible
line identification in the observed spectra. The lower curve in
Fig.~\ref{exomodels} shows a solar-composition model with the same \Teff\ and
\logg\ but with a different redshift, namely $z$=0.24. The observed lines at
$\lambda$=13.0/13.75~\AA\ could stem from \ion{Fe}{XXIV} (n=2--3,
$\lambda_0$=10.6--11.4~\AA ). The \ion{O}{VIII}~Ly$_\alpha$ line is located at
23.6~\AA\ in the $z$=0.24 redshifted model and a weak feature is seen in the
observed spectrum at this location. We would like to stress that our analysis is
in a very preliminary stage. But our models clearly indicate that a re-analysis
of the X-ray burst spectra of \exo\ is necessary.

\begin{figure}
\centering
  \includegraphics[width=0.7\columnwidth]{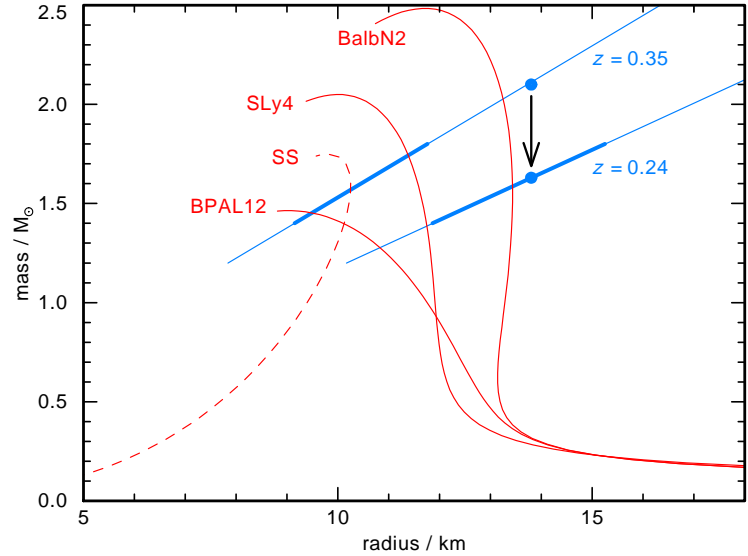}
\caption{Allowed values for $M$ and $R$ of \exo\ for redshifts $z$=0.24
and $z$=0.35 (straight lines; thick portions of the graphs denote the mass
range 1.4--1.8~\msun) compared to various theoretical $M$--$R$ relations
\citep{Hae06}.
The thick dot on the
$z$=0.35 line denotes the minimum $M$ and $R$ derived by \cite{ozel06}. The
arrow indicates the shift of this result when we assume $z$=0.24.}
\label{eos}
\end{figure}

Fig.~\ref{eos} shows the allowed values for NS mass and radius for redshifts
$z$=0.24 and $z$=0.35 compared to various theoretical mass-radius
relations. While $z$=0.35 gives radii of $R$=9--12~km for a mass-range of
$M$=1.4--1.8~\msun, our redshift $z$=0.24 gives larger radii, namely 12--15~km,
which corresponds to stiff equations-of-state and excludes mass-radius
relations based on exotic matter. This result is in line with a recent study of
\exo\ by \citet{ozel06} using additional observational constraints. With
$z$=0.35 she derives minimum values of mass and radius, $M$\,$\geq$2.10~\msun\
and $R$\,$\geq$13.8~km. A reduction of the redshift to $z$=0.24 has a negligible
effect on her radius determination but the lower mass limit is reduced to
1.63~\msun.

{\bf Acknowledgements} TR is supported by BMBF/DESY (grant 05 AC6VTB) and VS by
DFG (grant We 1312/35-1) and Russian FBR (grant 05-02-17744).

\end{document}